\documentclass[lettersize,journal]{IEEEtran}
\usepackage{amsmath,amsfonts}
\usepackage{algorithmic}
\usepackage{algorithm}
\usepackage{array}
\usepackage[caption=false,font=normalsize,labelfont=sf,textfont=sf]{subfig}
\usepackage{textcomp}
\usepackage{stfloats}
\usepackage{url}
\usepackage{verbatim}
\usepackage{cite}
\usepackage[pdftex]{graphicx}
\usepackage{booktabs}
\usepackage{cite}
\usepackage{multicol}
\usepackage{multirow}
\usepackage{tabularx}
\usepackage{threeparttable}
\usepackage{xcolor}
\usepackage{pifont}
\usepackage{makecell}
\begin{document}

\title{Inter-LPCM: Learning-based Inter-Frame Predictive Coding for LiDAR Point Cloud Compression }
\author{Chang Sun, Hui Yuan,~\IEEEmembership{Senior Member,~IEEE,} Shiqi Jiang, Chongzhen Tian, Guanghui Zhang, Raouf Hamzaoui,~\IEEEmembership{Senior Member,~IEEE} 
        % <-this % stops a space
          \vspace{-0.9cm}
\thanks{This work was supported in part by the National Natural Science Foundation of China under Grants 62222110, 62172259, 62571303, and the High-end Foreign Experts Recruitment Plan of Chinese Ministry of Science and Technology under Grant G2023150003L, the Taishan Scholar Project of Shandong Province (tsqn202103001), the Natural Science Foundation of Shandong Province under Grant ZR2022ZD38, the Key Technology Research and Development Program of Shandong Province 2024CXGC010212.
\textit{(Corresponding author: Hui Yuan.)}}% <-this % stops a space
\thanks{Chang Sun, Hui Yuan, Shiqi Jiang and Chongzhen Tian are with the School of Control Science and Engineering, Shandong University, Jinan 250061, China, and also with the Key Laboratory of Machine Intelligence and System Control, Ministry of Education, Jinan 250061, China (e-mail: huiyuan@sdu.edu.cn).}
\thanks{Guanghui Zhang is with the School of Computer Science and Technology, Shandong University, Qingdao, 266237, China.}
\thanks{Raouf Hamzaoui is with the School of Engineering and Sustainable Development, De Montfort University, LE1 9BH Leicester, U.K.}
}

% The paper headers
\markboth{Journal of \LaTeX\ Class Files,~Vol.~14, No.~8, August~2021}%
{Shell \MakeLowercase{\textit{et al.}}: A Sample Article Using IEEEtran.cls for IEEE Journals}

\maketitle

\begin{abstract}
Because LiDAR sensors acquire point clouds with a fixed angular resolution, the resulting data can be systematically parameterized and efficiently compressed in the spherical coordinate system. Traditional spherical coordinate-based point cloud compression methods have shown strong rate-distortion (RD) performance, with the predictive geometry coding (PredGeom) method in the geometry-based point cloud compression (G-PCC) standard being a prominent example. While PredGeom includes an inter-frame prediction mode, it relies on a simple linear model, which limits its ability to capture complex motion patterns or structural dependencies. On the other hand, existing learning-based compression methods in the spherical domain do not exploit inter-frame correlations to reduce geometry redundancy. 
To address these limitations, we propose a learning-based inter-frame predictive coding method (Inter-LPCM). For azimuth prediction, we use a delta coding strategy based on the predefined angular resolution. To improve compression for radii, we introduce an inter-frame radius predictive (Inter-RP) model that estimates the current point’s radius using neighboring points from both the current frame and the registered reference frame. In addition, we design a lightweight attention-based prediction (LAEP) model to predict elevation angles by capturing long-range geometric correlations across different coordinates.
For quantization, we propose an RD-optimized method to select the quantization steps in the spherical coordinate system. For entropy coding, we design distinct models for each spherical coordinate component. These models are adapted to the statistical priors of each coordinate, which enables more accurate probability estimation. Experimental results show that Inter-LPCM, in its best RD configuration, achieved a D1-PSNR BD-rate reduction of $26.1\%$ compared with the G-PCC lossless octree-based coding mode on \textit{SemanticKITTI}, and $8.3\%$ compared with the inter-frame prediction mode of PredGeom on \textit{Ford}, using the latest G-PCC test model TMC13 v31.0. Our source code is publicly available at https://github.com/SDUChangSun/Inter-LPCM.
\end{abstract}


\section{Introduction}
\IEEEPARstart{L}{iDAR} sensors measure distances by emitting laser beams and detecting their reflections. They are widely used in autonomous driving  \cite{1}, robotics \cite{2}, geographic information systems \cite{3}, etc. As shown in Fig. 1, LiDAR uses multiple laser emitters with different elevation angles to scan the environment horizontally at a predefined azimuth angular resolution. Each laser return produces a single point, which is defined by its azimuth angle $\phi$, elevation angle $\theta$, and measured distance $r$. These spherical coordinates are then converted into Cartesian coordinates to generate a LiDAR point cloud (LPC).

\begin{figure}[!t]
\centering
  \includegraphics[width=9cm]{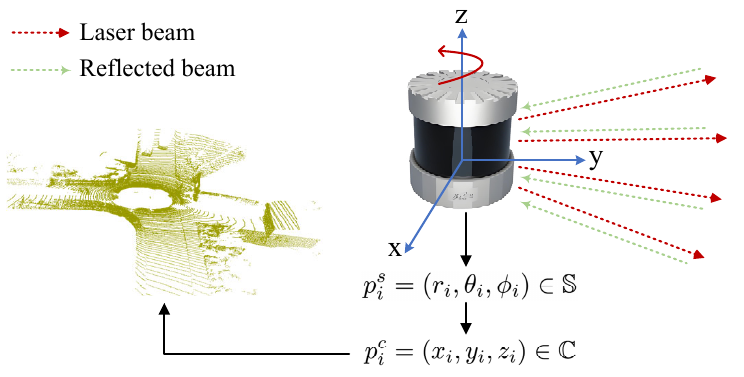} 
  \caption{Acquisition of LPCs. First, LiDAR uses multiple laser emitters with different elevation angles $\theta$ to scan in the horizontal direction at a predefined azimuth angular resolution. Then, the distance $r$ between the LiDAR and the object is calculated based on the time of flight of the reflected beams. Finally, the azimuth angle $\phi$, elevation angle $\theta$, and the measured distance $r$ of each laser beam are converted into the Cartesian coordinate system to form a LiDAR point cloud (LPC).}
\label{fig1}
  \vspace{-0.6cm}
\end{figure}

Due to the huge data volume of LPCs, efficient compression is essential to reduce storage and transmission costs. However, points in LPCs are irregularly distributed in the Cartesian coordinate system, making efficient compression challenging. Fortunately, since LPCs are acquired with predefined angular resolutions, they can be regularly organized and effectively compressed in the spherical coordinate system. 

Recently, spherical coordinate-based methods have shown strong rate-distortion (RD) performance for LPC compression. Luo et al. \cite{4} introduced two such methods: spherical-coordinate-based EHEM (SCP-EHEM) and spherical-coordinate-based OctAttention (SCP-OctAttention). These approaches convert an LPC into an octree within a spherical coordinate system. They then use either EHEM \cite{5} or OctAttention \cite{6} to estimate voxel occupancy probabilities, which enables efficient entropy coding. A key limitation of these methods, however, is their failure to explicitly account for the inherent angular resolution of LiDAR sensors.

The predictive geometry coding (PredGeom) method \cite{7}\cite{8} in the geometry-based point cloud compression (G-PCC) standard \cite{9} first converts all points into the spherical coordinate system. Points with the same laser ID are then ordered by increasing azimuth angle and connected to form a predictive tree. The geometry of each point is sequentially predicted along this tree starting from the root, and the resulting residuals are subsequently entropy encoded. 
LiDAR sensors have a fixed azimuth resolution, which allows the azimuth angle of the current point to be predicted from that of the previous point. In addition, points originating from the same laser emitter often share similar elevation angles. The G-PCC PredGeom module exploits these inherent angular characteristics to efficiently encode azimuth and elevation angles, achieving greater compression efficiency than octree-based methods \cite{10} for LPCs. However, PredGeom relies on a simple linear model based on neighboring points, neglecting long-range geometric correlations.

In \cite{11}, we proposed a learning-based predictive coding method (LPCM) that represents LPCs as predictive trees for efficient compression. LPCM uses a delta coding (DC) scheme to compress azimuth angles and radii, and introduces a learning-based predictive module to reduce the error between actual and ideal elevation angles. We also analyzed the impact of distortions in the radius, elevation, and azimuth components on overall geometry distortion and proposed a quantization step (Qs) selection strategy. Despite its effectiveness, LPCM has several limitations. First, it relies on a simple linear model that considers only neighboring points to predict the radius. Since the radius of a point depends solely on the distance between the LiDAR sensor and the object, it lacks prior information and thus requires significantly more bits than the elevation and azimuth components. Second, the Qs selection strategy in LPCM is based exclusively on peak signal-to-noise ratio (PSNR), which does not always align with optimal RD performance. Third, LPCM uses a general context-adaptive binary arithmetic coder (CABAC) \cite{12} to compress residuals, without incorporating domain-specific optimizations adapted to spherical coordinates.

Building on LPCM, we propose three major enhancements: incorporating inter-frame information to enhance coordinate prediction, selecting Qs values through RD optimization, and designing entropy encoders adapted to the statistical characteristics of spherical coordinate residuals. To improve radius compression, we analyze points from different laser emitters to partition the LPC into ground and non-ground subsets. Based on this partition, we develop an inter-frame radius prediction (Inter-RP) model that estimates each point’s radius using neighboring points from both the current and registered reference frames. For elevation prediction, we introduce a lightweight attention-based elevation prediction (LAEP) model that captures long-range geometric correlations across coordinates. To determine Qs values for each residual component, we propose an RD-optimized selection method that uses Bjøntegaard Delta rate (BD-rate) as the evaluation metric. Finally, for entropy coding, we build distinct entropy models for each spherical coordinate to enable accurate probability estimation. Experimental results show that Inter-LPCM, in its best RD configuration, achieved a D1-PSNR BD-rate reduction of $26.1\%$ compared with the G-PCC lossless octree-based coding mode on \textit{SemanticKITTI}, and $8.3\%$ compared with the inter-frame prediction mode of PredGeom on \textit{Ford}, using the latest G-PCC test model TMC13 v31.0 \cite{13}. Inter-LPCM also achieved D1-PSNR BD-rate reductions of $6.4\%$ and $4.3\%$ compared to the current state-of-the-art (LPCM) on \textit{SemanticKITTI} and \textit{Ford}, respectively. Furthermore, the reconstructed point clouds yield the best performance in vehicle detection tasks. The main contributions of this work are summarized as follows.

\begin{enumerate}
\item {We propose Inter-LPCM, a learning-based inter-frame predictive coding method that combines efficient radius prediction, azimuth prediction, RD-optimized Qs selection, and tailored entropy models for predicting the residuals of each component. To the best of our knowledge, this is the first learning-based inter-frame coding method in the spherical coordinate system.}
\item {For radius prediction, we partition the points into ground and non-ground subsets and propose an Inter-RP model, which predicts each point’s radius using neighboring points from both the current frame and the registered reference frame. In addition, we propose an LAEP model that predicts elevation angles by capturing long-range geometric correlations across different coordinates.}
\item {For residual quantization, we propose an RD-optimized Qs selection method based on the differential evolution (DE) algorithm.}
\item {For entropy coding, we design distinct entropy models for each spherical coordinate component based on their statistical priors, which enables accurate probability estimation and improves coding efficiency.}
\end{enumerate}

The remainder of this paper is organized as follows. Section 2 reviews related work. Section 3 and 4 describe the proposed method. Section 5 presents and analyzes the experimental results. Finally, Section 6 concludes the paper and suggests future work.

\section{Related work}

In this section, we review traditional geometry compression methods and learning-based geometry compression methods.

\subsection{Traditional Geometry Compression Methods}
The G-PCC standard \cite{9}\cite{14} proposed by the Moving Picture Experts Group (MPEG) is one of the most representative traditional method for geometry compression. G-PCC defines three geometry compression approaches: trisoup-based coding \cite{15}, octree-based coding\cite{10}, and PredGeom \cite{7}\cite{8}. Among them, trisoup-based coding relies on a dense and continuous surface to efficiently compress dense point clouds. However, as LPCs are typically sparse, trisoup is less effective for their compression. The octree-based method first voxelizes the point cloud and represents the voxelized data using an octree structure. Then, the occupancy information of each node in the octree is entropy encoded with OBUF \cite{16}. However, as the octree-based method does not exploit the angular resolution of LPCs to reduce geometry redundancy, its compression efficiency is limited. PredGeom converts a point cloud into the spherical coordinate system and connects points with the same laser ID in order of their azimuth angles to construct a predictive tree. The spherical coordinates of each point are then sequentially predicted starting from the root point. Finally the residuals are entropy encoded. PredGeom also provides an inter-prediction mode \cite{17}. Specifically, the radius of the current point is predicted from the point in the reference frame that has the same laser ID and the closest azimuth angle. Although PredGeom outperforms octree-based method and most learning-based methods \cite{11}\cite{18}, it relies only on a simple linear predictive model based on neighboring points in the predictive tree or the reference frame, while neglecting long-range geometry correlation.

In addition to G-PCC, many other effective traditional geometry compression methods have been proposed. Rather than encoding discrete point positions, Krivokuća et al. \cite{19} proposed a volumetric function-based approach that uses B-spline wavelet bases to represent geometry as level sets. Tzamarias et al. \cite{20} proposed a fast run-length intra-frame lossless geometry compression method that operates directly in the voxel domain, achieving an average speedup of 1.8 times over G-PCC. Li et al. \cite{21} proposed a hierarchical prior-based super-resolution framework for point cloud geometry compression, which enhances the reconstruction quality of G-PCC by transmitting a content-dependent prior to the decoder. However, these methods are primarily designed for dense point clouds. For LPCs compression, Yu et al. \cite{22} observed that LPCs have irregular distributions in the spherical coordinate system and proposed a trigonometric function-based prediction model to regularize them for efficient compression. However, this method does not account for LiDAR acquisition errors or environmental influences. 

  \vspace{-0.3cm}
\subsection{Learning-based Geometry Compression Methods}
Recently, learning-based methods have become widely used for point cloud compression \cite{23}. These methods can be broadly categorized into two main approaches: autoencoder-based methods and autoregressive entropy models.

Autoencoder-based methods are commonly used for lossy compression. Huang et al. \cite{24} introduced an autoencoder to compress point cloud geometry. The encoder extracts a latent representation from the point cloud, the representation is entropy coded, and the decoder reconstructs the point cloud from the decoded representation. Quach et al. \cite{25} proposed an RD-joint loss function that enables simultaneous optimization of bitrate and distortion. Later works \cite{26,27,28,29} integrated entropy models to estimate the probability distribution of latent representations more accurately, which improved entropy coding performance. Wu et al. \cite{30} developed a method that represents human point clouds with geometric priors and structural deviations. For inter-frame coding, Akhtar et al. \cite{31} designed a motion compensation network that operates in the feature space. This network predicts the latent representation of the current frame from the previous one, and a learning-based entropy model then compresses the residuals. However, these methods are primarily designed for dense point clouds and are less effective for LPCs due to their inherent sparsity.

Autoregressive entropy model-based methods first represent point clouds as voxels or octrees. The entropy models estimate occupancy probabilities for voxels (0 for an empty voxel, 1 for an occupied voxel) or for octree nodes (with 255 possible occupancy states). These probabilities are then used in entropy coding. The distortion in the reconstructed point cloud depends only on the precision of the voxelization or the octree representation. For voxel-based entropy models, Nguyen et al. \cite{32}\cite{33} proposed entropy models that estimate the occupancy probability of each voxel based on encoded voxels. However, due to the voxel-by-voxel operation, the time complexity is very high. To reduce the coding time, MS-VoxelDNN \cite{34} organizes voxels within a multiscale architecture that models voxel occupancy in a coarse-to-fine order. Fan et al. \cite{35} proposed a method that builds a hierarchical entropy model with latent variables as context. The method also introduces a residual coding scheme based on latent variables and applies soft addition to improve compression efficiency. SparsePCGC \cite{36} exploits both inter-scale and intra-scale correlations to achieve efficient compression. Wang et al. \cite{37} proposed a unified, low-complexity compression method called Unicorn, which predicts occupancy probabilities for entropy coding from encoded frames. However, because LPCs are sparse, voxelization generates a large number of empty voxels, which leads to inefficient coding. You et al. \cite{38} proposed RENO, a real-time neural codec for 3D LiDAR point cloud compression, which uses multiscale sparse tensor representations and one-shot sparse occupancy codes to achieve efficient coding with low computational complexity. However, all the aforementioned methods do not leverage the angular characteristic.

For octree-based entropy models, OctSqueeze \cite{39} introduced a multilayer perceptron (MLP)-based entropy model that predicts the probability distribution of node occupancy. Fu et al. \cite{6} proposed an attention-based entropy model called OctAttention, which uses large-scale context to estimate occupancy probabilities. However, OctAttention encodes octree nodes one by one, which results in high complexity. EHEM \cite{5} later adopted a grouped context structure that improves coding efficiency. In our previous work \cite{40}\cite{41}, we observed that entropy models based on cross-entropy loss and one-hot encoding capture only differences in the positions of sub-nodes and fail to capture differences in the number of occupied sub-nodes. To address this problem, we introduced a sub-node number prediction module that assists the entropy model in estimating probability distributions. Biswas et al. \cite{42} proposed MuSCLE, an inter-frame entropy model that uses ancestor nodes and previously encoded frames to predict occupancy probabilities. Song et al. \cite{43} proposed an inter-frame entropy model for dynamic point cloud compression, which exploits a geometry-aware graph feature extraction module to capture large-scale spatial-temporal context. Luo et al. \cite{44} propose a misalignment-aware compression framework that mitigates both macroscopic and microscopic misalignment to address the intrinsic misalignment issue in dynamic LiDAR point cloud sequences. Wang et al. \cite{45,46,47,48} proposed a series of octree-based point cloud geometry compression methods that progressively enhance spatial and channel-wise context modeling. Specifically, OctGLP-Net \cite{45} captures spatial correlations among octree nodes and enables fine-grained feature interactions across spatial and channel dimensions through global–local perception. TopNet \cite{46} introduces a Transformer-based octree occupancy prediction network to improve local–global dependency modeling. Building on cross-dimensional interactions, GCFI-Net \cite{47} further integrates global spatial context, global–local channel information, and hierarchical features from ancestor and sibling nodes to enhance octree node representations. To address redundancy and noise introduced by efficient attention mechanisms, they proposed the Adaptive Sparse Representation Learning (ASRL) framework \cite{48}, which suppresses noisy interactions from irrelevant nodes in octree-based compression. However, all the aforementioned learning-based methods compress point clouds in the Cartesian coordinate system. To address this, SCP-EHEM and SCP-OctAttention \cite{4} represent LPCs as spherical-coordinate-based octrees and apply EHEM \cite{5} or OctAttention \cite{6} for probability estimation. Similarly, Cui et al. \cite{49} introduced a graph-driven attention-based entropy model that organizes point clouds with a multi-level spherical octree to improve reconstruction quality. While these works \cite{4}, \cite{49} represent LPCs in the spherical coordinate system, they neglect the angular resolution of LiDAR.

In our previous work \cite{11}, we first represented LPCs as predictive trees and applied a DC-based method to compress azimuth angles and radii. We then proposed a learning-based predictive module to minimize the error between actual and ideal elevation angles. Since LPCM uses the angular resolution, it compresses elevation and azimuth angles efficiently. However, similar to PredGeom, LPCM still relies on a simple linear model to predict the radius of the current point. LPCM also uses a general-purpose CABAC \cite{12} to encode the residuals, which limits compression efficiency.

\begin{figure*}[!t]\centering
  \includegraphics[width=18cm]{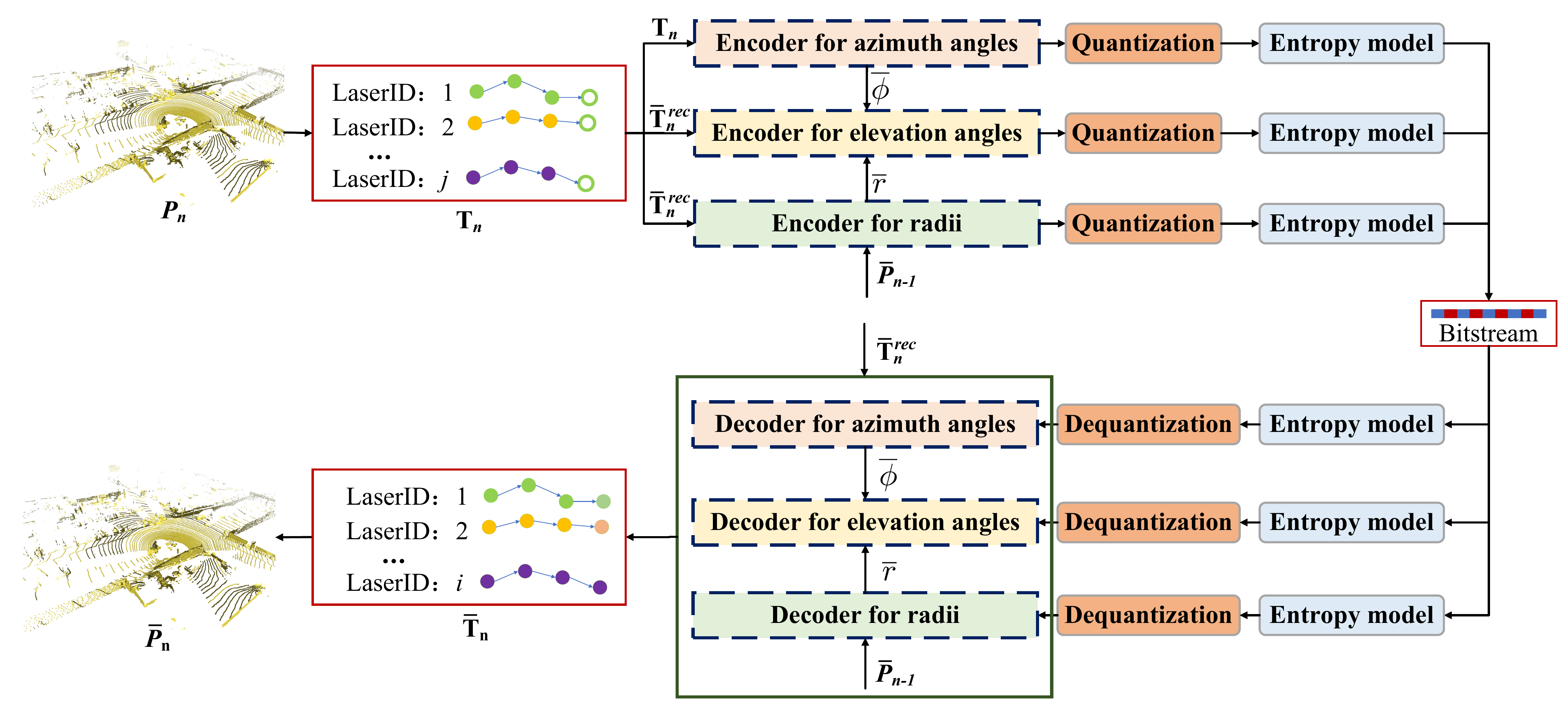}
  \caption{Overall architecture of the proposed Inter-LPCM. First, the current point cloud $\boldsymbol{P}_{n}$ is represented as predictive trees $\mathbf{T}_{n}$. Then, each component of the current point is compressed using its corresponding predictor and entropy model. For the azimuth angle $\phi$ and elevation angle $\theta$, the reconstructed predictive tree $\overline{\mathbf{T}}_{n}^{rec}$ serves as context. For the radius $r$, the reconstructed reference point cloud $\boldsymbol{\overline{P}}_{n-1}$ is registered to the current point cloud $\boldsymbol{P}_{n}$ and represented as predictive trees $\overline{\mathbf{T}}_{n-1}$. Both $\overline{\mathbf{T}}_{n}^{rec}$ and $\boldsymbol{\overline{P}}_{n-1}$ serve as context for compressing the radius.}
  \label{fig4}
\vspace{-0.5cm} 
\end{figure*}

\section{Proposed Method}
The architecture of the proposed Inter-LPCM is shown in Fig. 2. First, an LPC is represented as multiple predictive trees based on the laser ID. For azimuth angles, we use the azimuth angular resolution of the LiDAR and apply the DC between the current point $\boldsymbol{p}_i$ and its preceding point $\boldsymbol{p}_{i-1}$ in the predictive tree to represent the azimuth angle of $\boldsymbol{p}_i$. For elevation angles, the proposed LAEP model exploits the correlations among neighboring points and the dependency between radius and elevation angle to predict the elevation angle of the current point. Usually, the bit cost of compressing radii is significantly higher than that for azimuth and elevation angles\cite{7}\cite{11}. To reduce the coding cost of radii, we introduce temporal information. Specifically, we first determine whether the current frame $\boldsymbol{P}_{n}$ should be intra-coded (I-frame) by calculating its similarity to the preceding frame $\boldsymbol{P}_{n-1}$. For I-frames, we use a DC-based method to compress the radii. For predictive-coded frames (P-frames), we compute and encode a transformation matrix (T-matrix) between the current frame and the reference frame (R-frame) to register the R-frame to the current frame. Then, the radius of the current point is predicted using the proposed Inter-RP model based on its neighboring points in the current frame and the registered R-frame. Finally, the residuals of radii, elevation angles and azimuth angles are quantized using the Qs values determined by the proposed Qs selection algorithm and then compressed with the proposed entropy models. 

\subsection{Predictive Tree Construction}
Considering the variations in LiDAR calibration parameters and the arrangement of points in LiDAR point cloud files, we use the predictive tree construction method proposed in \cite{11} to construct the predictive trees. We first convert the LPC $\boldsymbol{P}_{n}$ into a spherical coordinate system and group the points based on their laser ID. A point $\boldsymbol{p}_i$ in Cartesian coordinates $\boldsymbol{p}_i^{\text{Cartesian}}(x, y, z)$ can be represented in spherical coordinates as

\begin{equation}
\boldsymbol{p}_i^{\text{Spherical}}(r, \theta, \phi, id),
\end{equation}
where $r$ is the radius, $\theta$ is the elevation angle, $\phi$ is the azimuth angle, and $id$ is the laser ID. Within each group, the points are sorted in ascending order of azimuth angle to construct a predictive tree $T_{n,id}$. During encoding, the spherical coordinates of each point are encoded sequentially according to the order in $T_{n,id}$. For each point, the azimuth angle is encoded first, followed by the radius, and finally the elevation angle.

\subsection{Compression of azimuth angles} 
The azimuth angles of LPC points are generally close to integer multiples of the LIDAR's angular resolution. To efficiently encode these angles, we introduce a preset unit angle $\phi_{\text{unit}}$, which defines the discretization step for azimuth angles: 
\begin{equation}
\phi_{\text{unit}} = \frac{\phi_{\text{ar}}}{q_{\phi}},
\end{equation}
where $\phi_{\text{ar}}$ is the LiDAR's azimuth angular resolution, and $q_{\phi}$ is an integer Qs that controls the granularity of the representation. Each azimuth angle $\phi_i$ of a point $\boldsymbol{p}_i$ is then represented by an integer slope $s_i$ defined as $s_i =\lfloor\frac{\phi_i}{ \phi_{\text{unit}}}\rceil$. The residual between successive azimuth slopes can be expressed as
\begin{equation}
\Delta s_{i} = s_i-s_{i-1}.
\end{equation}
As shown in Fig. 3, $\Delta s_{i}$ is generally greater than or equal to $\phi_{\text{ar}}$, which results in an asymmetric probability distribution of $\Delta s_{i}$. Therefore, we propose an entropy model that uses a skew-normal distribution \cite{50} to capture the probability distribution of $\Delta s_{i}$.

\begin{figure}[!t]
\centering
  \includegraphics[width=8cm]{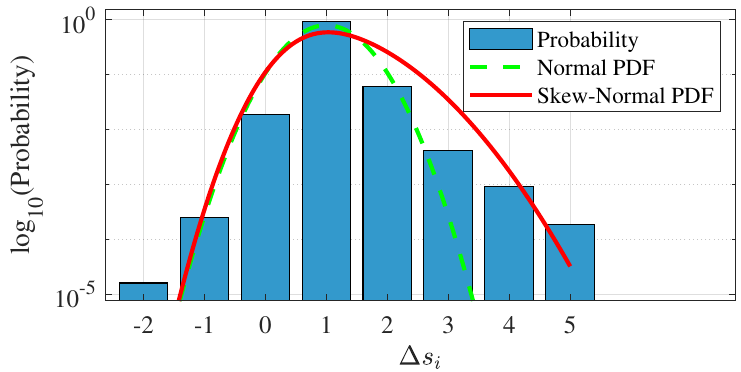}
  \vspace{-0.3cm} 
  \caption{Histogram of the probability distribution of $\Delta s_{i}$ (with $q_{\phi}=1$), overlaid with the probability density functions of both a normal distribution and a skew-normal distribution. The y-axis shows the logarithm (base 10) of the probability. ‘Normal PDF’ refers to the probability density function of a normal distribution, while ‘Skew-normal PDF’ refers to that of a skew-normal distribution. The histogram shows the probability distribution of $\Delta s_{i}$.}
\label{fig2}
\vspace{-0.1cm} 
\end{figure}

\begin{figure}[!t]
\centering
  \includegraphics[width=8cm]{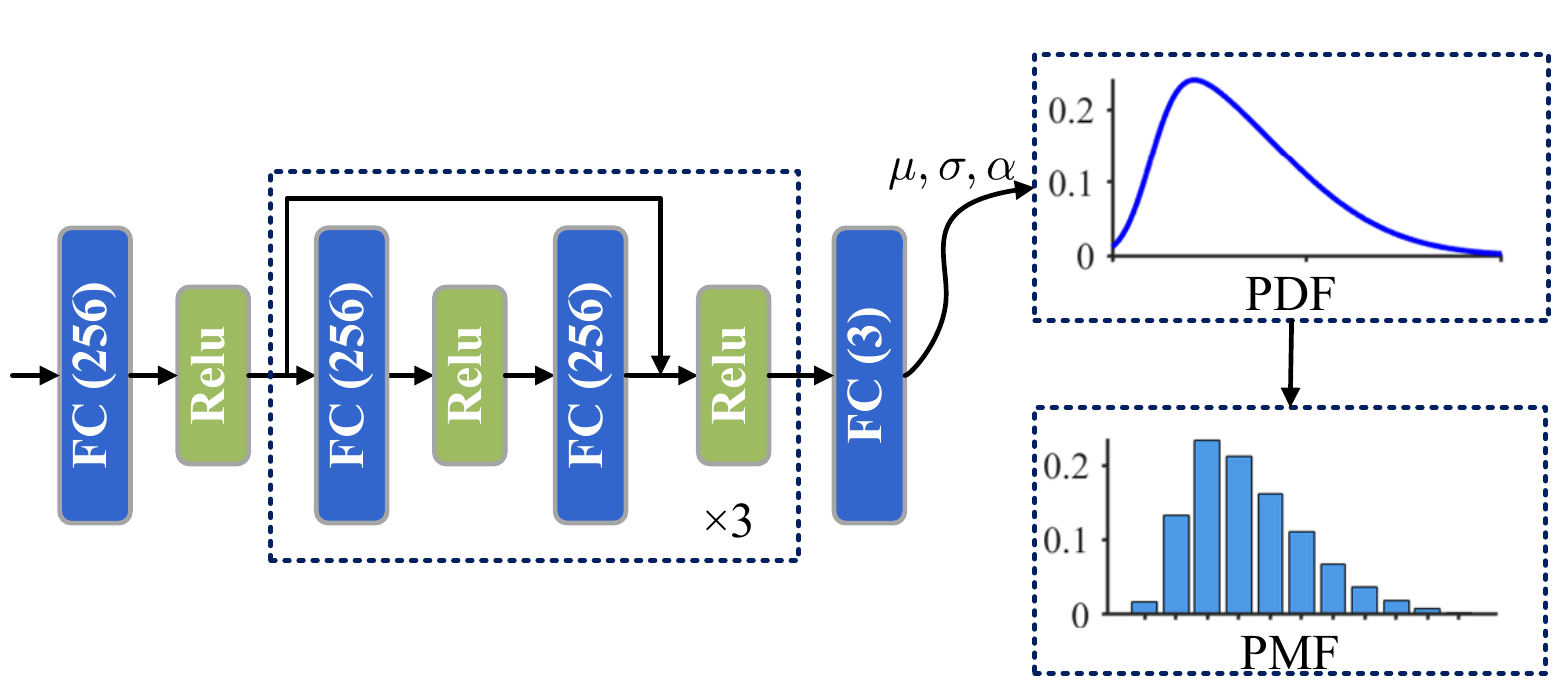}
  \vspace{-0.1cm} 
  \caption{Architecture of the proposed entropy model. The encoded residuals are fed into a stack of fully connected (FC) layers with ReLU activations. The final FC layer outputs the location parameter $\mu_i$, the scale parameter $\sigma_i$, and the skewness parameter $\alpha_i$ to model a probability density function (PDF) of a skew-normal distribution. This PDF is then converted into a probability mass function (PMF) for entropy coding. Entropy models with a similar structure are also used for compressing the radius and elevation angle. In these cases, the final FC layer outputs only $\mu_i$ and $\sigma_i$ to define a normal distribution.}
\label{fig2}
\vspace{-0.1cm} 
\end{figure}

The architecture of the proposed entropy model  for azimuth angles is illustrated in Fig. 4. First, the context formed by the encoded residuals $\{\Delta s_{i-50},\Delta s_{i-49},...,\Delta s_{i-1}\}$ is fed into an entropy model composed of stacked residual networks. Then, the entropy model outputs the location $\mu_i$, the scale $\sigma_i$, and the skewness $\alpha_i$ to define a skew-normal distribution that models the probability distribution of $\Delta s_{i}$. The skew-normal distribution is expressed as

\begin{equation}
f_{\text{SN}}(\Delta s_{i}; \mu_i, \sigma_i, \alpha_i) = \frac{2}{\sigma_i} \, \varphi\left(\frac{\Delta s_{i} - \mu_i}{\sigma_i}\right) \, \Phi\left(\alpha_i \cdot \frac{\Delta s_{i} - \mu_i}{\sigma_i}\right)
\end{equation}
where $\varphi(\cdot)$ is the standard normal probability density function (PDF), defined as
\begin{equation}
\varphi(x) = \frac{1}{\sqrt{2\pi}} e^{-\frac{1}{2} x^2 },
\end{equation}
and $\Phi(\cdot)$ is the standard normal cumulative distribution function (CDF), defined as
\begin{equation}
\Phi(x) = \int_{-\infty}^{x} \frac{1}{\sqrt{2\pi}} e^{ -\frac{1}{2} t^2 } dt.
\end{equation}
To encode $\Delta s_{i}$, $f_{\text{SN}}(\Delta s_{i}; \mu_i, \sigma_i, \alpha_i)$ must be converted into a discrete probability mass function (PMF)
\begin{equation}
p(\Delta s_{i}) = \int_{\Delta s_{i}-0.5}^{\Delta s_{i}+0.5}f_{\text{SN}}(t; \mu_i, \sigma_i, \alpha_i)dt.
\end{equation}

At the decoder, we reconstruct $s_i$ using the reconstructed $s_{i-1}$ and the decoded $\Delta s_{i}$. The reconstructed azimuth angle $\overline{\phi}_{i}$ is then obtained as

\begin{equation}
\overline{\phi}_{i} = {\phi_{\text{unit}}} \times {s_i} .
\end{equation}

To train the entropy model, we use the negative log-likelihood loss, defined as
\begin{equation}
\mathcal{L}_{\text{NLL}} = -\sum_{i=1}^{N} \log p(\Delta s_i),
\end{equation}
where $N$ is the number of points in the LPC $\boldsymbol{P}_{n}$.

\subsection{Compression of radii} 
In spherical coordinate prediction, the azimuth and elevation angles can be predicted using LiDAR priors. However, the radius of a point depends only on the distance between the LiDAR and the object, making its compression more challenging. To enhance the prediction accuracy, we incorporate inter-frame information. As shown in Fig. 5, we compress the radius through five stages: partitioning, registration, prediction, quantization, and entropy coding. In partitioning, both the current frame $\boldsymbol{P}_{n}$ and the reconstructed R-frame $\boldsymbol{\overline{P}}_{n-1}$ are partitioned into an upper part $\boldsymbol{P}^{\text{upper}}$ and a lower part $\boldsymbol{P}^{\text{lower}}$ based on the radius variance of predictive tree. In the registration, we first determine whether the current frame $\boldsymbol{P}_{n}$ is an I-frame or a P-frame by evaluating the similarity between $\boldsymbol{P}_n^{\text{upper}}$ and $\boldsymbol{\overline{P}}_{n-1}^{\text{upper}}$. For I-frames, $\boldsymbol{P}_{n}$ is represented using predictive trees $\mathbf{T}^I_n$. For P-frames, $\boldsymbol{\overline{P}}_{n-1}^{\text{upper}}$ is first registered to $\boldsymbol{P}_n^{\text{upper}}$ by calculating the T-matrix between them. The radius of each point in I-frames and in $\boldsymbol{P}^{\text{lower}}$ of P-frames is predicted by using a DC-based method, whereas the radius in $\boldsymbol{P}^{\text{upper}}$ of P-frames is predicted using the proposed Inter-RP model. Then, the  residuals are quantized and entropy encoded using the proposed entropy model. 

\begin{figure*}[!t]\centering
  \includegraphics[width=18cm]{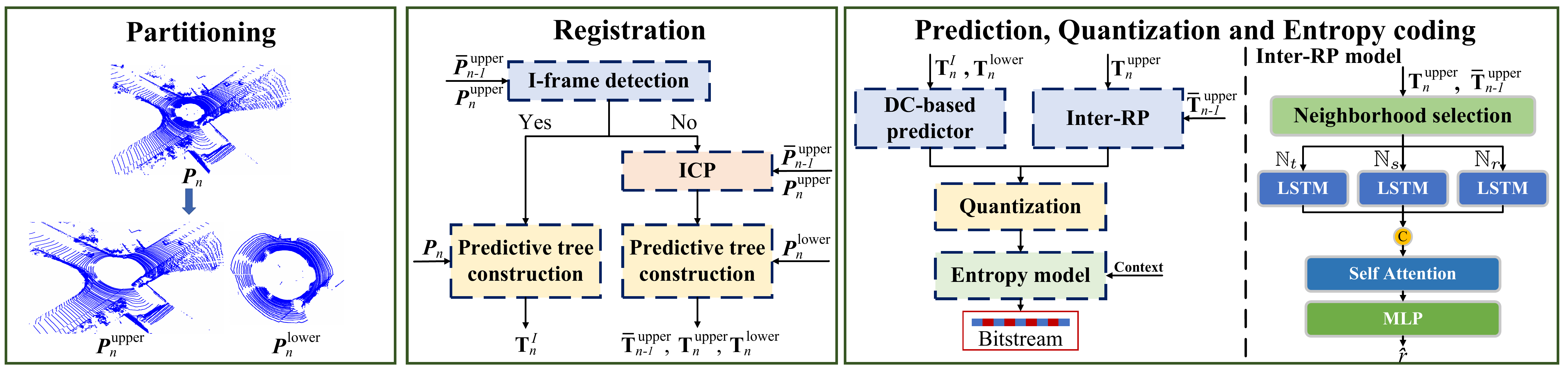}
  \caption{Compression pipeline for radii. In the partitioning stage, both the current frame $\boldsymbol{P}_{n}$ and the reconstructed reference frame $\boldsymbol{\overline{P}}_{n-1}$ are partitioned into an upper part and a lower part based on the radius variance of predictive trees. In the registration stage, we first determine whether the current frame $\boldsymbol{P}_{n}$ is an I-frame or a P-frame by evaluating the similarity between $\boldsymbol{P}_n^\text{upper}$ and $\boldsymbol{\overline{P}}_{n-1}^\text{upper}$ where the superscript “$\text{upper}$” denotes the upper part of the corresponding frame. For I-frames, $\boldsymbol{P}_{n}$ is directly represented as a predictive tree $\mathbf{T}^I_n$. For P-frames, $\boldsymbol{\overline{P}}_{n-1}^\text{upper}$ is registered to $\boldsymbol{P}_n^\text{upper}$. Then, the registered $\boldsymbol{\overline{P}}_{n-1}^\text{upper}$, $\boldsymbol{P}_n^\text{upper}$ and $\boldsymbol{P}_n^\text{lower}$ are represented as $\overline{\mathbf{T}}_{n-1}^\text{upper}$, $\mathbf{T}_n^\text{upper}$ and $\mathbf{T}_n^\text{lower}$, respectively, where the superscript “$\text{lower}$” denotes the lower part of the corresponding frame. In the prediction stage, the points in $\mathbf{T}^I$ and $\mathbf{T}^\text{lower}$ of the P-frames are predicted using a DC-based predictor, while the points in $\mathbf{T}_n^\text{upper}$ of the P-frames are predicted using the proposed Inter-RP model. The Inter-RP model selects neighborhoods of current point from $\mathbf{T}_n^\text{upper}$ and $\overline{\mathbf{T}}_{n-1}^\text{upper}$ to form the temporal neighborhood $\mathbb{N}_t$, spatial neighborhood $\mathbb{N}_s$ and residual neighborhood $\mathbb{N}_r$. Each neighborhood is processed by an LSTM, and their outputs are fused using a self-attention layer. The predicted radius $\hat{r}$ is then output by an MLP. In the quantization and entropy coding stage, the residuals are quantized and entropy encoded using the proposed entropy model. The context input to the entropy model consists of the encoded residuals of the radii.}
  \label{fig4}
\vspace{-0.1cm} 
\end{figure*}

\textbf{Partitioning.} We partition the current frame $\boldsymbol{P}_{n}$ into an upper part $\boldsymbol{P}^\text{upper}_n$ and a lower part $\boldsymbol{P}^\text{lower}_n$ based on the radius variance of predictive trees and encode them separately. This design is motivated by the LiDAR acquisition mechanism. As shown in Fig.1, LiDAR acquires point clouds using multiple laser emitters with fixed elevation angles. Predictive trees associated with lower elevation angles mainly contain dense ground points with relatively similar radii, while trees associated with higher elevation angles include farther points and therefore show larger radius variation. In addition, the upper region typically contains objects (e.g., trees and buildings) that are more stable in world coordinates, whereas the lower region is dominated by ground points. As the LiDAR platform moves, the current frame and the reference frame may cover different ground areas, which reduces overlap and is less favorable for reliable registration \cite{51}.

\begin{figure}[!t]
\centering
  \includegraphics[width=7cm]{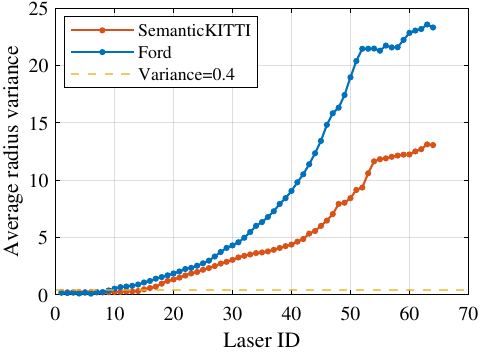}
  \vspace{-0.1cm}
  \caption{Radius variance of predictive trees with different laser IDs}
  \label{fig8}
  \vspace{-0.3cm}
\end{figure}

Therefore, as shown in Fig. 5, we use the proposed radius-variance-based partitioning method to partition $\boldsymbol{P}_{n}$ into $\boldsymbol{P}^\text{upper}_n$ and $\boldsymbol{P}^\text{lower}_n$. For each predictive tree $T_{n,id}$, we compute the variance of the radii of its points, $v_{n,id}$. As shown in Fig. 6, we then find the smallest index $t$ such that both $v_{n,t}$ and $v_{n,t+1}$ exceed a predefined threshold $\tau$ (set to $0.4$ in the main setting). The frame is then partitioned as
\begin{equation}
P_n^{\mathrm{lower}}=\bigcup_{id\le t} T_{n,id}, \qquad
P_n^{\mathrm{upper}}=P_n\setminus P_n^{\mathrm{lower}},
\end{equation}
The split index $t$ is encoded into the bitstream to ensure deterministic decoding. 

\textbf{Registration.} After partitioning, we determine whether the current frame $\boldsymbol{P}_{n}$ is an I-frame by computing the PSNR between the upper parts of the current and previous frames, denoted as $\text{PSNR}(\boldsymbol{P}_n^\text{upper}, \boldsymbol{\overline{P}}_{n-1}^\text{upper})$. If $\text{PSNR}(\boldsymbol{P}_n^\text{upper},\boldsymbol{\overline{P}}_{n-1}^\text{upper})$ is less than a predefined threshold, $\boldsymbol{P}_{n}$ is classified as an I-frame Otherwise, $\boldsymbol{P}_{n}$ is classified as a P-frame. The rationale is that a low PSNR indicates a large difference between consecutive frames, which usually corresponds to a significant scene change or unreliable inter-frame prediction. In such cases, coding the current frame as an I-frame is more appropriate. Conversely, when the PSNR is high, the current and previous frames are sufficiently similar, and inter-frame coding can be used. For the I-frame, $\boldsymbol{P}_{n}$ is directly represented as predictive trees $\mathbf{T}_n$. For the P-frame, $\boldsymbol{\overline{P}}_{n-1}^\text{upper}$ is registered to $\boldsymbol{P}_n^\text{upper}$ using the iterative closest point (ICP) algorithm. The transformation matrices obtained by the ICP algorithm are encoded into the bitstream for motion compensation at the decoder. We also clarify that ICP is used only to improve motion-compensated prediction. Even if ICP is inaccurate, it does not affect the correctness of encoding or decoding, because the residuals are still explicitly encoded.
Subsequently, the registered $\boldsymbol{\overline{P}}_{n-1}^\text{upper}$, which is obtained from the R-frame, together with $\boldsymbol{P}_n^\text{upper}$ and $\boldsymbol{P}_n^\text{lower}$ from the P-frame, are respectively represented as $\overline{\mathbf{T}}_{n-1}^\text{upper}$, $\mathbf{T}_n^\text{upper}$, and $\mathbf{T}_n^\text{lower}$.

\textbf{Prediction.} For I-frames and $\mathbf{T}_n^{\text{lower}}$ (from P-frames), the radius $r_i$ of the point $\boldsymbol{p}_i$ is predicted from the reconstructed radius $\overline{r}_{i-1}$ of its preceding neighbor $\boldsymbol{p}_{i-1}$ in the predictive tree, giving
\begin{equation}
\hat{r}_i = \overline{r}_{i-1}.
\end{equation}

The radii of points in $\mathbf{T}_n^{\text{upper}}$ (from P-frames) are predicted using the proposed Inter-RP model. Specifically, the Inter-RP model first selects the neighborhood of $\boldsymbol{p}_i$ from $\mathbf{\overline{T}}_{n-1}^{\text{upper}}$ and reconstructed part of $\mathbf{T}_n^{\text{upper}}$ to form the spatial neighborhood $\mathbb{N}_s$, the temporal neighborhood $\mathbb{N}_t$ and the residual neighborhood $\mathbb{N}_r$. To construct $\mathbb{N}_s$, we select a predefined number of reconstructed neighboring points within the predictive tree that have the closest azimuth angle to $\boldsymbol{p}_i$. The residual neighborhood $\mathbb{N}_r$ consists of the encoded residuals of points in $\mathbb{N}_s$. To construct $\mathbb{N}_t$, we first select the predictive trees from $\overline{\mathbf{T}}_{n-1}^\text{upper}$ whose laser ID satisfy $id \in [id_{i}-1,id_{i}+1]$, where $id_{i}$ is the laser ID of the current point $\boldsymbol{p}_i$. From each selected predictive tree, we then choose the subset of $k$ points whose azimuth angles are closest to that of $\boldsymbol{p}_i$. Finally, we take the union of these subsets to form $\mathbb{N}_t$.

The $\mathbb{N}_s$, $\mathbb{N}_t$ and $\mathbb{N}_r$ are individually processed by long short-term memory (LSTM) networks \cite{52}, and the resulting features are fused through a self-attention layer. Finally, the predicted radius $\hat{r}_i$ is obtained using a multilayer perceptron (MLP). This process is formulated as
\begin{equation}
\hat{r}_i = f(\mathbb{N}_s, \mathbb{N}_t, \mathbb{N}_r \mid \psi),
\end{equation}
where $\psi$ denotes the learnable parameters of the proposed Inter-RP model. 

\textbf{Quantization and entropy coding.} The residual of the radius $res_{r,i}$ and its quantized version $\widetilde{res}_{r,i}$ are defined as
\begin{equation}
res_{r,i} = r_i - \hat{r}_i,
\end{equation}
\begin{equation}
\widetilde{res}_{r,i} = \left\lfloor{res_{r,i}}\times{q_r} \right\rceil,
\end{equation}
where $q_r$ is the Qs of the radius. Finally, $\widetilde{res}_{r,i}$ is encoded using the proposed entropy model. The only difference between this entropy model and the one used for the azimuth angle is that this entropy model assumes a normal distribution rather than a skew-normal distribution. The model outputs the predicted mean $\mu_i$ and predicted standard deviation $\sigma_i$  of the PDF as

\begin{equation}
f_{\text{N}}(\widetilde{res}_{r,i}; \mu_i, \sigma_i)  = \frac{1}{\sigma_i\sqrt{2\pi}} e^{-\frac{(\widetilde{res}_{r,i}-\mu_i)^2}{2\sigma_i^2} }.
\end{equation}
To encode $\widetilde{res}_{r,i}$, we convert $f_{\text{N}}(\widetilde{res}_{r,i}; \mu_i, \sigma_i)$ into a PMF
\begin{equation}
p(\widetilde{res}_{r,i}) = \int_{\widetilde{res}_{r,i}-0.5}^{\widetilde{res}_{r,i}+0.5}f_{\text{N}}(t; \mu_i, \sigma_i)dt.
\end{equation}

The Inter-RP model is trained with the mean squared error (MSE) loss between the actual radius $r_{i}$ and the predicted radius $\hat{r}_{i}$:
\begin{equation}
\mathcal{L}_{\text{MSE}}=\frac1{N}\sum_{i=1}^{N}{\left\|r_{i}- \hat{r}_{i}\right\|_2^2}, 
\end{equation}
where $N$ is the total number of points.

At the decoder, the reconstructed radius $\overline{r}_{i}$ is recovered as 
\begin{equation}
\overline{r}_{i} = \hat{r}_i + \frac{\widetilde{res}_{r,i}}{q_r}.
\end{equation}
\subsection{Compression of elevation angle} 
The elevation angles of points captured by the same laser emitter are ideally identical. However, due to internal calibration errors and acquisition noise of the LiDAR, these elevation angles often vary in practice. Points captured by the same laser emitter typically share strong spatial correlation, particularly between points with similar azimuth angles. On the other hand, a point's radius and elevation angle are negatively correlated \cite{11}. 

To capture both the spatial correlation among neighboring points and the correlation between radius and elevation angle, we propose the LAEP model (Fig. 7). The input to the LAEP model consists of a predefined number of reconstructed neighboring points of current point $\boldsymbol{p}_i$ from the predictive tree, along with a virtual current point $\boldsymbol{p}_i^v(\overline{r}_i , \overline{\theta}_{i-1}, \overline{\phi}_i, id_i)$, which contains the reconstructed radius $\overline{r}_i$ and the reconstructed azimuth angle $\overline{\phi}_i$ used to predict the elevation angle $\overline{\theta}_{i}$. 
The neighboring points are first fed into a three-layer LSTM network to capture the spatial correlation between the current point and its previously reconstructed neighboring points. 

The motivation for using LSTM \cite{52} in both the Inter-RP and LAEP models is that geometry coding along the predictive tree naturally forms an ordered sequence (defined by laser ID and increasing azimuth). Each point is predicted conditioned on previously reconstructed points, so the model needs to capture sequential dependencies along the tree. LSTM updates a hidden state step by step, which makes it well suited for modeling local spatial correlations between neighboring points as well as longer-range dependencies along the sequence.

The features output by the LSTM are composed of the hidden states corresponding to each input point, where each hidden state encodes information about both the radius and azimuth angle. To capture the correlation between the radius and azimuth angle, we apply a self-attention layer to adaptively weight each hidden state. Then, the weighted hidden states are aggregated through an MLP. The resulting feature is added to the mean elevation angle $e^m$ of the current predictive tree, which serves as a direct current component, to output the predicted elevation angle $\hat{\theta}_{i}$. The residual $res_{\theta,i}$ and quantized residual $\widetilde{res}_{\theta,i}$ can then be expressed as
\begin{equation}
res_{\theta,i} = \theta_i - \hat{\theta}_i,
\end{equation}
\begin{equation}
\widetilde{res}_{\theta,i} = \left\lfloor {res_{\theta,i}}\times{q_\theta} \right\rceil,
\end{equation}
where $q_\theta$ is the Qs for the elevation angle. Finally, $\widetilde{res}_{\theta,i}$ is encoded using the proposed normal entropy model. 

\begin{figure}[!t]
\centering
  \includegraphics[width=9cm]{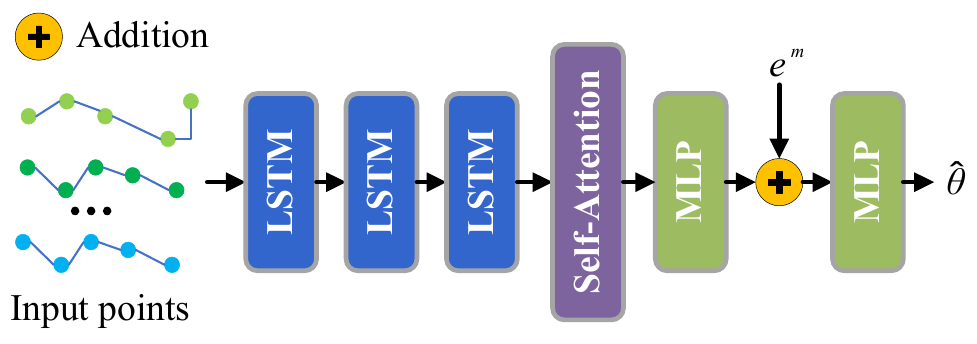}
  \vspace{-0.5cm} 
  \caption{Architecture of the proposed LAEP model. The model consists of three LSTM layers, a self-attention layer, and two MLPs. Its input consists of the reconstructed spherical coordinates $(\overline{r}, \overline{\theta}, \overline{\phi}, id)$ of a preset number of neighboring points in the predictive tree, along with the virtual current point $\boldsymbol{p}_i^v(\overline{r}_i , \overline{\theta}_{i-1}, \overline{\phi}_i, id_i)$. The output of the first MLP is added to the mean elevation angle $\theta^m$ of the current predictive tree. The second MLP outputs the predicted elevation angle $\hat{\theta}_{i}$.}
\label{fig3}
\end{figure}

The proposed LAEP model is trained using the MSE loss between the actual elevation angle $\theta_{i}$ and the predicted elevation angle $\hat{\theta}_{i}$:
\begin{equation}
\mathcal{L}_{\text{MSE}}=\frac1{N}\sum_{i=1}^{N}{\left\|\theta_{i}- \hat{\theta}_{i}\right\|_2^2}, 
\end{equation}

At the decoder, the reconstructed elevation angle $\overline{\theta}_{i}$ is expressed as
\begin{equation}
\overline{\theta}_{i} = \hat{\theta}_i + \frac{\widetilde{res}_{\theta,i}}{ q_\theta}.
\end{equation}

\section{RD-Optimized Quantization Step Selection}
In our previous work \cite{11}, we showed that each axis in the spherical coordinate system represents a different physical quantity, and that radius reconstruction error has a much larger impact on overall distortion than errors in azimuth and elevation. However, most existing encoding-parameter selection methods \cite{53}\cite{54} model the relationship between Qs and bitrate in the Cartesian coordinate system and do not account for inter-axis differences. As a result, they are not well suited for selecting Qs in spherical coordinates. Hou et al. \cite{55} observed that increasing the quantization difference between angular axes and the radius axis can improve RD performance, but this conclusion was based on empirical observations and did not provide an explicit model of the relationship between Qs and RD behavior. In \cite{11}, we proposed an evolutionary Qs selection method, but it used only PSNR as the optimization objective. To address this limitation, we formulate the problem as minimizing the sum of BD-rates under constraints on the average bitrates over multiple point clouds. For clarity, we illustrate the selection procedure using six target bitrates as an example:

\begin{equation}
 \begin{gathered}
\min_{\boldsymbol{Q}} \quad\sum_{n=1}^N \text{BDRate}(\boldsymbol{Q}, \boldsymbol{Q}^{\text{ref}},\boldsymbol{P}_n) \\
\mathbf{s.t.}\  R_i^{\text{min}}\leq \frac{1}{N}\sum_{n=1}^N R(\boldsymbol{q}_i, \boldsymbol{P}_n) \leq R_i^{\text{max}},\ i\in \{1,2,...,6\}
\end{gathered}
\end{equation}
where $\{\boldsymbol{P}_n\}_{n=1}^{N}$ denotes the set of $N$ point clouds used for Qs selection.
$\text{BDRate}(\boldsymbol{Q}, \boldsymbol{Q}^{\text{ref}},\boldsymbol{P}_n)$ is the BD-rate between the two RD curves on $\boldsymbol{P}_n$ generated by $\boldsymbol{Q}$ and $\boldsymbol{Q}^{\text{ref}}$ using the six operating bitrates.
Here $\boldsymbol{Q} = (\boldsymbol{q}_1, \boldsymbol{q}_2, \dots, \boldsymbol{q}_6)$ is the set of quantization-step tuples corresponding to the six target bitrates, where
$\boldsymbol{q}_i = (q_{i,\phi}, q_{i,\theta}, q_{i,r})$ represents the quantization steps for azimuth, elevation, and radius at the $i$-th bitrate.
$\boldsymbol{Q}^{\text{ref}}$ is a manually selected reference set that achieves favorable RD performance.
Note that $\boldsymbol{Q}^{\text{ref}}$ is used only as a reference for BD-rate computation and does not affect the final optimization result.
$R_i^{\min}$ and $R_i^{\max}$ are the lower and upper bounds of the $i$-th target bitrate range, which can be set to suit specific requirements.

As the correlation between the distortion on each axis and the overall reconstruction error remains consistent \cite{11}, Qs values derived from a subset of point clouds tend to perform comparably on the entire dataset. To reduce complexity, the DE optimization algorithm is applied to four sequences from the \textit{SemanticKITTI} dataset, with five consecutive frames from each sequence (4x5=20 LPCs) to obtain a set of Qs values. Specifically, we first randomly initialize the candidate solutions, called the population in the DE algorithm 
\begin{equation}
 \begin{gathered}
\mathbb{Q}(t) = (\boldsymbol{Q}_{1}(t),\boldsymbol{Q}_{2}(t),...,\boldsymbol{Q}_{K}(t)), \\
\end{gathered}
\end{equation}
where $\mathbb{Q}(t)$ represents the population after $t$ iterations (or generations), $\boldsymbol{Q}_{k}(t)=(\boldsymbol{q}_{1,k}(t), \boldsymbol{q}_{2,k}(t), \dots, \boldsymbol{q}_{6,k}(t))$ is the $k-th$ individual in $\mathbb{Q}(t)$, and $K$ is the population size. To ensure computational tractability of the DE optimization while covering Qs values that provide meaningful RD performance, we restrict the search ranges to
\begin{equation}
 \begin{gathered}
q_{\theta}, q_{r} \in \left\{x \,\middle|\, x \in \mathbb{Z}, \, 1 \leq x \leq 256 \right\}, \\
q_{\phi} \in \left\{ x \,\middle|\, x \in \mathbb{Z}, \, 1 \leq x \leq 16 \right\}.
\end{gathered}
\end{equation}
During the DE optimization, mutation and crossover are used to maintain population diversity. In the mutation step, the difference between two individuals is used to compute  the perturbation $\boldsymbol{B}_{k,l}(t)$
\begin{equation}
\boldsymbol{B}_{k,l}(t) = \boldsymbol{Q}_{k}(t) - \boldsymbol{Q}_{l}(t).
\end{equation}
Then, $\boldsymbol{B}_{k,l}(t)$ is scaled and added to another individual $\boldsymbol{Q}_{j}(t)$ to obtain the mutant individual $\boldsymbol{V}_{j}(t+1)$:
\begin{equation}
\boldsymbol{V}_{j}(t+1) = \boldsymbol{Q}_{j}(t) + \mu\boldsymbol{B}_{k,l}(t),
\end{equation}
where $\mu$ is the scaling factor. In the crossover step, $\boldsymbol{V}_{j}(t+1)$ and $\boldsymbol{Q}_{j}(t)$ exchange elements according to the crossover rate $CR$ to form the trial individual $\boldsymbol{U}_{j}(t+1)$.

\begin{table}[t]
    \centering
    \begin{threeparttable}
        \caption{Quantization steps}
        \label{tab:1}
        \begin{tabular}{ccccccc}
            \toprule
             & $r01$ & $r02$ & $r03$ & $r04$ & $r05$ & $r06$  \\
            \midrule
            $q_\phi$    & 1  & 2  & 2  & 4  & 8   & 8      \\
            $q_\theta$  & 2  & 3  & 4  & 15  & 33   & 61    \\
            $q_r$       & 9   & 18   & 34 & 66 & 121  & 172  \\
            \bottomrule
        \end{tabular}
    \end{threeparttable}
    \vspace{-0.3cm}
\end{table}

\begin{table}[t]
    \centering
    \begin{threeparttable}
        \caption{Bitrate constraints for SemanticKITTI and Ford}
        \label{tab:constraints}
        \begin{tabular}{ccc}
            \toprule
            & SemanticKITTI & Ford \\
            \midrule
            $r01$ & [2.8, 3.0] & [3.0, 3.2] \\
            $r02$ & [3.5, 3.8] & [4.0, 4.3] \\
            $r03$ & [4.4, 4.8] & [5.0, 5.3] \\
            $r04$ & [5.2, 5.6] & [5.8, 6.2] \\
            $r05$ & [6.8, 7.2] & [7.2, 7.6] \\
            $r06$ & [8.8, 9.2] & [9.0, 9.4] \\
            \bottomrule
        \end{tabular}
    \end{threeparttable}
    \vspace{-0.3cm}
\end{table}

After mutation and crossover, we evaluate the fitness of $\boldsymbol{Q}_{j}(t)$ and $\boldsymbol{U}_{j}(t+1)$. If any $\boldsymbol{q}_{i,j}(t)$ in $\boldsymbol{Q}_{j}(t)$ or $\boldsymbol{U}_{j}(t+1)$ violates the bitrate constraint
\[
R_i^{\text{min}}\leq \frac{1}{N}\sum_{n=1}^N R(\boldsymbol{q}_{i,k}(t), \boldsymbol{P}_n) \leq R_i^{\text{max}},\ i\in \{1,2,...,6\}
\]
the fitness of that individual is set to $+\infty$. Otherwise, the fitness function is defined as
\begin{equation}
 \begin{gathered}
f(\boldsymbol{Q}_{j}(t)) = \sum_{n=1}^N \text{BDRate}(\boldsymbol{Q}_{j}(t), \boldsymbol{Q}^{\text{ref}},\boldsymbol{P}_n).
\end{gathered}
\end{equation}
The next generation is then selected according to fitness
\begin{align}
\boldsymbol{Q}_{j}(t+1) =
\begin{cases} 
\boldsymbol{U}_{j}(t+1), & if\  f(\boldsymbol{U}_{j}(t+1)) < f(\boldsymbol{Q}_{j}(t)), \\
\boldsymbol{Q}_{j}(t), & \text{otherwise}.
\end{cases}
\end{align}
After a predefined number of iterations, the best individual $Q^*$ in the final population is selected. This $Q^*$ is then used to compress all LPCs. Both $Q^*$ and the bitrate constraints are listed in Table~\ref{tab:1} and Table~\ref{tab:constraints}, respectively.

\section{Experimental results}
\subsection{Datasets}
We evaluated the proposed method on two benchmark LiDAR datasets: \textit{SemanticKITTI} \cite{56} and MPEG \textit{Ford} \cite{57}, both collected with a Velodyne HDL-64 sensor.
The \textit{SemanticKITTI} dataset contains 43,552 LPCs captured in diverse environments, including urban streets, rural roads, and highways. Following the default split, sequences 00–10 were used for training, and sequences 11–21 were used for testing. The \textit{Ford} dataset, recommended by the MPEG Common Test Conditions (CTC) \cite{58}, consists of three sequences totaling 4,500 frames. We used sequence 01 for training and sequences 02 and 03 for testing.

\subsection{Implementation Details}
We implemented the proposed Inter-LPCM in PyTorch. The Inter-RP model, LAEP model, and entropy models were trained for 50 epochs with a batch size of 256, using the Adam optimizer and a learning rate of 0.0001. 

For the Inter-RP model, the PSNR threshold for I-frame detection was manually set to 35. Each neighborhood ($\mathbb{N}_s$, $\mathbb{N}_t$ and $\mathbb{N}_r$) consisted of 50 points. For the LAEP model, the neighborhood included 49 points. For the entropy models, the context consisted of 50 encoded residuals. In the DE-based Qs selection, the number of iterations was 50, the population size $K$ was 20, the scale factor $\mu$ was 0.4, and the crossover rate $CR$ was 0.9. 

We improved the speed of the proposed method using a group-wise parallel strategy. The point cloud was divided into coding groups of 200 points, and up to 512 coding groups were processed in parallel in each encoding/decoding pass.

In addition to the full version of Inter-LPCM, we also provide a fast variant denoted as Inter-LPCM (fast), which replaces the proposed entropy models with a general-purpose CABAC \cite{12}. Specifically, the full Inter-LPCM is intended to show the best achievable RD performance of the proposed framework, whereas Inter-LPCM (fast) is designed as a more practical runtime-oriented variant.

All experiments were conducted on a server with an Intel(R) Xeon(R) Gold 6148 CPU and an NVIDIA GeForce RTX 4090 GPU with 24 GB of memory. 

\begin{figure*}[t]\centering
  \includegraphics[width=18.5cm]{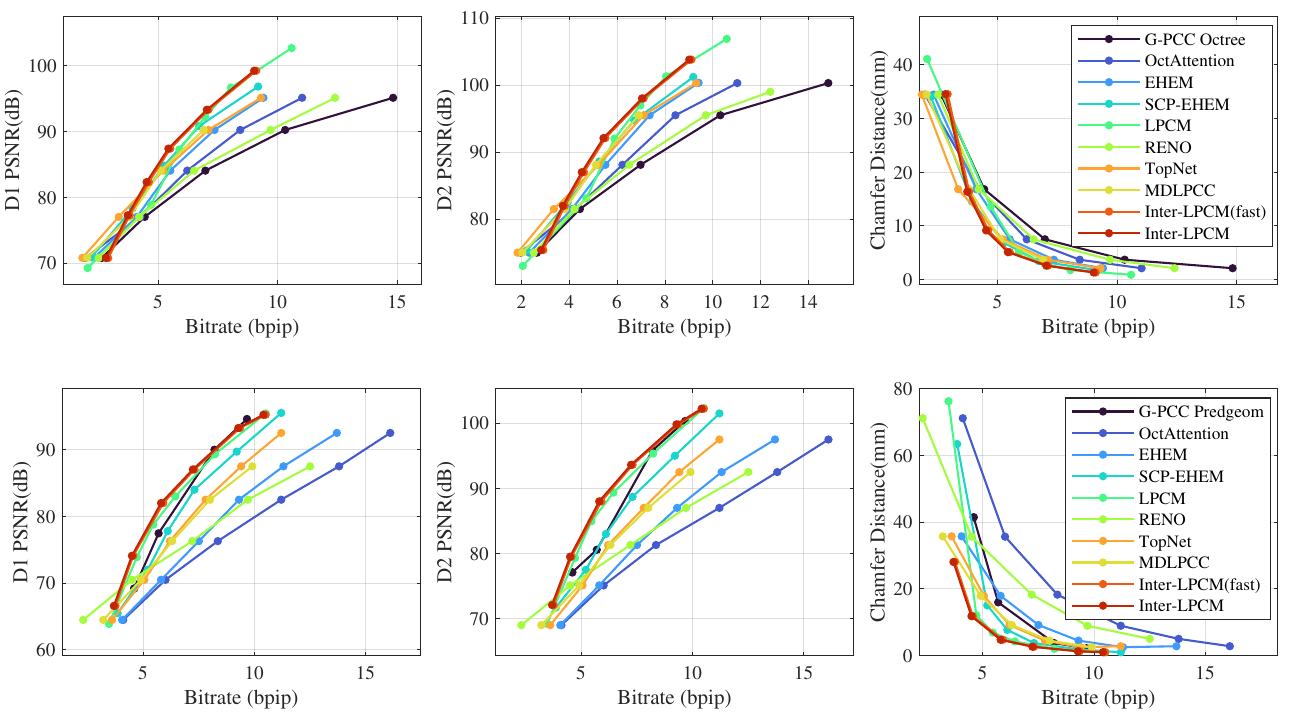}
  \vspace{-0.4cm}
  \caption{RD performance of the proposed method and the baselines on the \textit{SemanticKITTI} (top) and \textit{Ford} (bottom) datasets.}
  \label{fig4}
  \vspace{-0.4cm}
\end{figure*}

\begin{table*}
\centering
\caption{BD-Rates when comparing Inter-LPCM with other methods in terms of D1 PSNR, D2 PSNR, and Chamfer distance (CD)}
\label{tab:bd_rate}
\begin{tabular}{lccccccccc}
\toprule
\multirow{2}{*}{Method} & \multicolumn{3}{c}{SemanticKITTI} & \multicolumn{3}{c}{Ford} \\  
 &   D1 PSNR & D2 PSNR&CD&D1 PSNR &D2 PSNR & CD \\ \midrule
G-PCC &  -26.1\% & -26.9\% & -31.6\% & -8.3\% & -7.4\% & -8.9\%  \\
OctAttention &  -15.6\% & -16.8\% & -16.5\% & -41.3\% & -44.2\% & -40.2\% \\
RENO &  -23.1\% & -24.1\% & -25.5\% & -27.8\% & -31.8\% & -30.4\% \\
EHEM &  -9.8\% & -10.9\% & -10.0\% & -33.6\% & -36.5\% & -33.8\% \\
SCP-EHEM &  -5.3\% & -5.9\% & -5.6\% & -14.5\% & -16.6\% & -15.2\% \\
TopNet &  -2.0\% & -3.1\% & -2.9\% & -16.4\% & -17.8\% & -16.9\% \\
MDLPCC &  -2.4\% &-4.4\% & -3.3\% & -21.8\% & -25.1\%& -22.4\% \\
LPCM &  -6.4\% & -7.9\% & -7.5\% & -4.3\% & -4.5\% & -5.6\% \\
Inter-LPCM (fast) & -1.4\% & -1.4\% &-1.5\% & -1.7\% & -1.7\% & -1.7\%& \\\bottomrule
\end{tabular}
\vspace{-0.1cm}
\end{table*}

\vspace{-0.2cm}
\subsection{Baselines}
To evaluate the effectiveness of the proposed Inter-LPCM, we compared it with spherical coordinate-based methods LPCM \cite{11}, SCP-EHEM \cite{4}, the octree-based methods TopNet\cite{46}, EHEM \cite{6} OctAttention \cite{5}, dynamic point cloud compression method MDLPCC \cite{44} and the real-time compression method RENO \cite{38}. We also compared our method with the G-PCC reference software TMC13 v31.0 \cite{13}. On the \textit{Ford} dataset, G-PCC was configured as PredGeom with inter-frame mode as specified in CTC \cite{58}. Since G-PCC does not provide configuration files with calibration parameters for the \textit{SemanticKITTI} dataset, the LPCs in the \textit{SemanticKITTI} dataset were first represented as octrees with depths ranging from 10 to 15 and then compressed using the losslessly octree-based coding method.

\subsection{Evaluation Metrics}
The bitrate was measured in bits per input point (bpip). The distortion was measured using point-to-point PSNR (D1 PSNR), point-to-plane PSNR (D2 PSNR), and Chamfer distance (CD), which were computed with the $PC_{error}$ software \cite{59} developed by MPEG. To compare RD performance with the baselines, we computed the BD-rate using the implementation in \cite{60} with Akima interpolation.

Moreover, we further evaluated the proposed method on a downstream vehicle detection task using Voxel-R-CNN \cite{61} as the detector. Detection accuracy was assessed on the reconstructed LPCs using average precision (AP), following the standard protocol with a mean Intersection-over-Union (mIoU) threshold of 0.7, which is widely used in object detection benchmarks.

\begin{figure*}[!t]\centering
  \includegraphics[width=17cm]{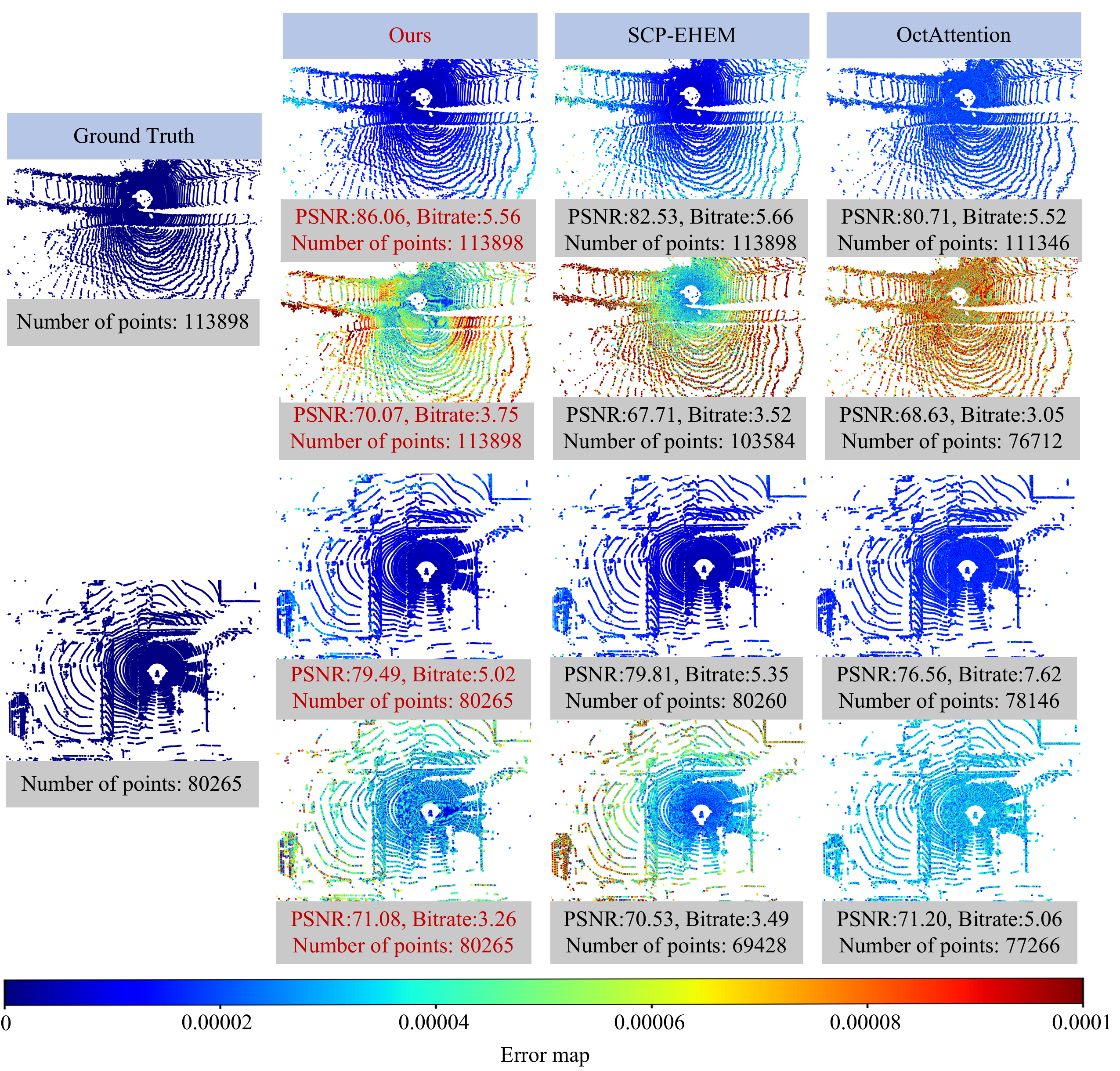}
  \vspace{-0.2cm}
  \caption{Visualization of point clouds reconstructed from OctAttention, SCP-EHEM, and the proposed Inter-LPCM on the \textit{SemanticKITTI} (top) and \textit{Ford} (bottom) datasets at two bitrates. Geometry coordinates are projected onto a plane orthogonal to the z-axis, and reconstruction errors are color-mapped to indicate their magnitudes.}
  \label{fig4}
  \vspace{-0.2cm}
\end{figure*}

\subsection{Experimental Results}
For the \textit{SemanticKITTI} dataset, Inter-LPCM outperformed all baselines at bitrates above 4.5 bpip (Fig. 8). Similarly, for the \textit{Ford} dataset, it surpassed all competing methods. Across both datasets, both the full version of Inter-LPCM and Inter-LPCM (fast) consistently achieved superior BD-rate performance compared with all baseline methods (Table~\ref{tab:bd_rate}).

The performance gain of our method on the \textit{Ford} dataset was greater than on the \textit{SemanticKITTI} dataset. This was because G-PCC provides configuration files for constructing predictive trees on the \textit{Ford} dataset, which allowed for more accurate tree structures. Furthermore, the performance improvement was particularly notable at higher bitrates. This was due to our method’s reliance on previously reconstructed points to predict the current points. At low bitrates, stronger quantization introduces larger distortion in reconstructed reference points, which weakens prediction accuracy and increases error propagation.

Fig. 9 shows point clouds reconstructed by OctAttention, SCP-EHEM, and the proposed Inter-LPCM. The geometric coordinates of each point cloud were projected onto a plane orthogonal to the z-axis, and reconstruction errors were color-mapped to indicate their magnitudes. The results show that Inter-LPCM achieved higher PSNR than the compared methods at similar bitrates on the \textit{Ford} dataset, while outperforming them only at higher bitrates on the \textit{SemanticKITTI} dataset. The distortion in OctAttention-reconstructed point clouds was uniform. In contrast, in point clouds reconstructed by Inter-LPCM and SCP-EHEM, points closer to the LiDAR sensor showed lower distortion than points farther away. This occurred because OctAttention used an octree structure, which produced uniform distortion, whereas Inter-LPCM applied globally unified Qs to the residuals of spherical coordinates. In application scenarios such as autonomous driving, the agent that perceives the environment via a LiDAR sensor should prioritize making decisions based on objects that are closer to it. Therefore, reducing distortion in the near field is reasonable.

\begin{table}[t]
\begin{center}
  \caption{Runtime, model size and GPU usage comparison}
  \label{tab:3}
  \begin{tabular}{ccccc}
    \toprule
     &Enc. &Dec. &Model size &GPU usage\\
     &(s/frame) &(s/frame) &(MB) &(MB)\\
    \midrule
    G-PCC & 1.03 & 0.66 &- &-\\
    OctAttention  &1.27 &306  &28.0 &836 \\
    TopNet  &0.38 &14.90  &13.5 &341 \\
    EHEM  &3.71 &3.92& 329 & 3218 \\
    SCP-EHEM  &3.77 &4.02  &329 & 3218 \\
    MDLPCC  &3.89 &4.75  &- & - \\
    LPCM  &3.35 &3.86& 1.35 & 325 \\
    Inter-LPCM (fast)  &5.23 &5.61 &7.20 &376  \\
    Inter-LPCM  &5.71 &86.7& 8.80 &2474 \\
  \bottomrule
\end{tabular}
\end{center}
\vspace{-0.5cm}
\end{table}

\begin{table}[t]
\centering
\caption{Accuracy  of vehicle detection at different bitrates.}
\label{tab:4}
\begin{tabular}{lcccc}
\toprule
\multirow{2}{*}{Method} & \multirow{2}{*}{Bitrate/D1-PSNR} &\multicolumn{3}{c}{Detection AP} \\
& & Easy & Mod &Hard\\
\midrule
Raw Data & - & 93.27 & 85.59 & 80.74 \\
\midrule
\multirow{3}{*}{G-PCC} & 2.63/ 70.8 & \textbf{85.71} & 78.73 & 75.98 \\

 & 4.43/ 77.0 & 86.94 & 78.31 & 76.17 \\

 & 6.97/ 84.0 & 90.04 & 79.83 & 77.09 \\
\midrule
\multirow{3}{*}{SCP-EHEM} & 2.27/ 70.1 & 85.40 & \textbf{79.03} & \textbf{76.33} \\

 & 4.12/ 78.3 & 88.28 & 79.13 & 76.67 \\

 & 6.71/ 90.8 & 92.18 & 84.08 & 78.93 \\
 \midrule
\multirow{3}{*}{LPCM} & 2.04/ 69.3 & 85.54 & 78.67 & 76.23 \\

 & 4.26/ 77.9 & 89.36 & 79.67 & 76.94 \\

 & 6.96/ 92.2 &93.20 & 85.32 & 80.18\\
\midrule
\multirow{3}{*}{Ours} & 2.26/ 68.2 & 85.35 & 78.47 & 76.06 \\

 & 4.15/ 78.5 & \textbf{89.90} & \textbf{80.16} & \textbf{77.28} \\

 & 6.75/ 92.8 & \textbf{93.25} & \textbf{85.37} & \textbf{80.28} \\
\bottomrule
\end{tabular}
\vspace{-0.1cm}
\end{table}

As shown in Table III, the full version of Inter-LPCM achieved the best RD performance. However, as shown in Table IV, its decoding complexity was high because of the proposed entropy models. In contrast, Inter-LPCM (fast) substantially reduced the decoding time compared with the full version, while still achieving BD-rate gains over all baselines (Table III). We also note that Inter-LPCM (fast) was slower than LPCM and G-PCC. Moreover, because Inter-LPCM (fast) does not use the proposed entropy models, it avoids the additional GPU memory required by group-wise parallel computation, making it more suitable for practical deployment than the full version.

\subsection{Downstream Applications}
Table~\ref{tab:4} presents vehicle detection performance. At high and medium bitrate settings, our method achieved higher reconstruction quality and correspondingly better downstream detection performance than the compared methods under equal or lower bitrate. At the low bitrate setting, although the reconstruction quality was lower than some baselines, our method still achieved competitive detection AP. A likely reason is that our method preserves the original number of points, which is beneficial for downstream object detection. In particular, at the lowest bitrate, LPCM achieved better reconstruction quality than Inter-LPCM, with a PSNR of 69.3 dB compared with 68.2 dB. This led to better downstream detection accuracy for LPCM at this bitrate. This is becasue LPCM includes a dedicated low-bitrate mode that combines the predictive-tree structure with an autoencoder, which is designed to improve reconstruction quality when very few bits are available. In contrast, Inter-LPCM follows the predictive-tree coding framework across all bitrates. Therefore, the lower detection accuracy of Inter-LPCM at the lowest bitrate is mainly caused by its lack of a dedicated autoencoder-based low-bitrate mode, rather than by a fundamental failure of the inter-frame architecture. In addition, detection performance on reconstructed point clouds becomes comparable to that on original point clouds only when PSNR exceeds about 90 dB, which indicates that high-quality reconstruction remains meaningful in practice.

\subsection{Ablation Study and Analysis}
We considered three variants of Inter-LPCM. First, to clearly demonstrate the contribution of each component, Inter-LPCM (fast) is denoted as \textbf{Abl-1} here. Next, based on \textbf{Abl-1}, we removed the Inter-RP model so that all point clouds were encoded as I-frames (\textbf{Abl-2}). Finally, based on \textbf{Abl-2}, we applied the Qs selection strategy proposed in [11] (\textbf{Abl-3}).

\begin{table}[t]
\begin{center}
  \caption{Ablation Study Results on BD-Rate and Runtime}
  \label{tab:5}
  \begin{tabular}{cccccccc}
    \toprule
     &\makecell{D1-PSNR}\\BD-rate &Enc. (s/frame) &Dec. (s/frame)\\
    \midrule
    \textbf{Abl-1}& -1.45\% & 5.23 &5.61\\
    \textbf{Abl-2}& -4.94\% & 3.74 &4.17 \\
    \textbf{Abl-3}& -6.08\% & - &- \\
  \bottomrule
\end{tabular}
\end{center}
\vspace{-0.3cm}
\end{table}

\begin{table}[t]
    \centering
    \begin{threeparttable}
        \caption{Selected Qs Values Using Different Numbers
of Point Clouds.}
        \label{tab:qs}
        \begin{tabular}{cccccccc}
            \toprule
            & \makecell{Number of\\ point clouds} & $r01$ & $r02$ & $r03$ & $r04$ & $r05$ & $r06$  \\
            \midrule
            
            \multirow{3}{*}{$q_\phi$}
            & 20 & 1 & 2 & 2 & 4 & 8 & 8 \\
            & 40 & 1 & 2 & 2 & 5 & 8 & 10 \\
            & 80 & 1 & 2 & 2 & 4 & 7 & 8 \\
            
            \midrule
            
            \multirow{3}{*}{$q_\theta$}
            & 20 & 2 & 3 & 4 & 15 & 33 & 61 \\
            & 40 & 2 & 4 & 4 & 16 & 35 & 63 \\
            & 80 & 2 & 3 & 4 & 15 & 32 & 59 \\
            
            \midrule
            
            \multirow{3}{*}{$q_r$}
            & 20 & 9 & 18 & 34 & 66 & 121 & 172 \\
            & 40 & 9 & 20 & 30 & 68 & 124 & 175 \\
            & 80 & 9 & 17 & 32 & 64 & 119 & 171 \\
            
            \bottomrule
        \end{tabular}
    \end{threeparttable}
    \vspace{-0.3cm}
\end{table}

\begin{figure}[t]
\centering
  \includegraphics[width=7cm]{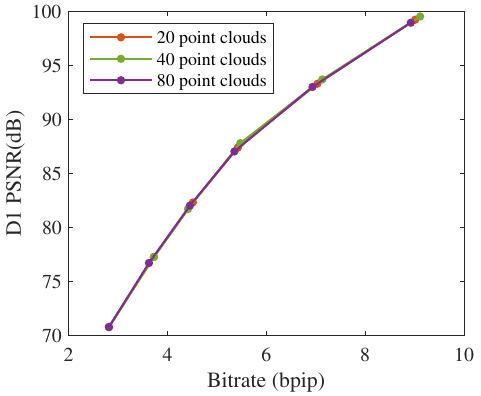}
  \vspace{-0.1cm} 
  \caption{Rate–D1 PSNR curves of selecting Qs values using different numbers of point clouds.}
\label{fig3}
\vspace{-0.1cm}
\end{figure} 

\begin{figure}[!t]
\centering
  \includegraphics[width=7cm]{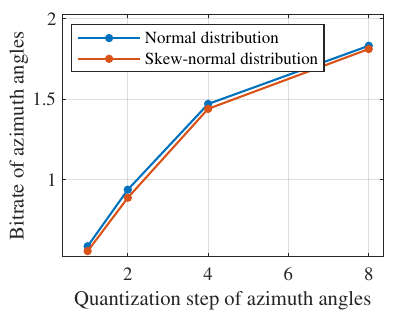}
  \vspace{-0.1cm} 
  \caption{Bitrates for compressing azimuth angles using the normal distribution and the skew-normal distribution.}
\label{fig3}
\vspace{-0.1cm}
\end{figure}

\begin{table}[t]
\begin{center}
\caption{Ablation Study Results on BD-Rate for Radius Coding}
\label{tab:6}
\begin{tabular}{cccc}
\toprule
Partitioning & Registration & I-frame detection &
\makecell{D1-PSNR\\BD-rate} \\
\midrule
\ding{55} & \ding{51} & \ding{51} & -0.48\% \\
\ding{55} & \ding{55} & \ding{51} & -1.81\% \\
\ding{55} & \ding{55} & \ding{55} & -2.29\% \\
\bottomrule
\end{tabular}
\end{center}
\vspace{-0.1cm}
\end{table}

\begin{figure}[t]
\centering
  \includegraphics[width=7cm]{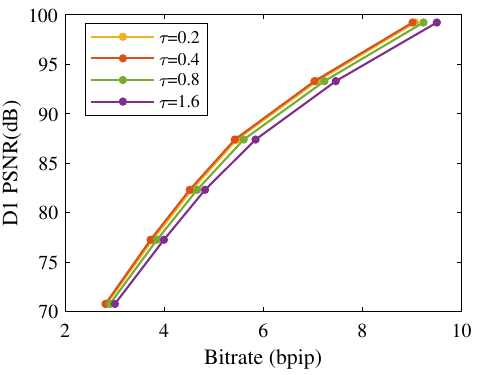}
  \vspace{-0.1cm} 
  \caption{Rate–D1 PSNR curves obtained by forming $\boldsymbol{P}^{\text{lower}}$ with thresholds $\tau = 0.2,0.4,0.8$ and $1.6$.}
\label{fig3}
\vspace{-0.1cm}
\end{figure} 

As shown in Table~\ref{tab:5}, Inter-LPCM outperformed \textbf{Abl-1} due to the more accurate probability distribution of residuals provided by the proposed entropy models. Nevertheless, \textbf{Abl-1} still achieves competitive RD performance, while significantly improving the decoding speed, demonstrating the proposed entropy model can be replaced with a faster entropy coder to achieve a better runtime-efficiency trade-off. Inter-LPCM also outperformed \textbf{Abl-2}, demonstrating the effectiveness of the proposed Inter-RP model. Compared with \textbf{Abl-3}, Inter-LPCM achieved further improvement, highlighting the contribution of the RD-optimized Qs selection method.

We added an ablation study to analyze the impact of the number of point clouds used for Qs selection. As shown in Table~\ref{tab:qs}, increasing the number of point clouds used for Qs selection did not significantly change the selected Qs values. Furthermore, as shown in Fig. 10, the corresponding RD curves remained very close, with no significant RD variation across different subset sizes. These results indicate that the selected Qs values are relatively stable and are not overly sensitive to the size of the optimization subset.

We compared the skew-normal distribution with the normal distribution. As shown in Fig. 11, the skew-normal distribution reduced the bitrate for compressing azimuth angles. This is because the azimuth angles of points within a predictive tree are theoretically monotonically increasing, and their increments are generally greater than or equal to the LiDAR’s azimuth resolution. Consequently, the skew-normal distribution better models the probability distribution of azimuth angles, resulting in a lower bitrate.

We also evaluated the effectiveness of each component in the encoder of Radius. As shown in Table~\ref{tab:6}, removing the partitioning stage leads to degraded RD performance. This degradation arises from variations in ground points induced by LiDAR motion, which in turn affect registration accuracy. Moreover, when the R-frame is not registered to the current frame, RD performance drops even further, since registration enables the Inter-RP model to identify similar regions within the R-frame. Finally, instead of adaptively determining whether the current frame should be encoded as an I-frame based on the PSNR between the R-frame and the current frame, we insert an I-frame every five frames. Under this fixed-interval strategy, the rate–distortion performance suffers additional deterioration, as the R-frame may fail to provide effective inter-frame information when it is dissimilar to the current frame.

We further evaluated threshold sensitivity with $\tau\in\{0.2,0.4,0.8,1.6\}$. Fig. 12 shows that $\tau=0.4$ yielded the best RD performance. This was because $\tau=0.4$ most accurately identified predictive trees mainly composed of ground points, which enabled more reliable registration between $\overline{P}_n^{\mathrm{upper}}$ and $P_n^{\mathrm{upper}}$ \cite{51}.

\section{Conclusion}
We presented Inter-LPCM, a learning-based inter-frame predictive coding method for LiDAR point clouds. The method exploits both temporal and spatial correlations to achieve efficient compression. The proposed Inter-RP model predicts the radius of each point using neighbors from the current frame and the registered R-frame. For quantization, differential evolution is used to select the quantization steps in spherical coordinates. For entropy coding, we introduced models for each component of the spherical coordinates, enabling accurate probability estimation and efficient coding. Experimental results showed that Inter-LPCM outperformed existing methods in rate-distortion performance and achieved superior results in a vehicle detection task. In future work, we plan to improve encoding and decoding speed through layer-wise coding.

\newpage

\vfill

\end{document}